\documentclass[twocolumn,prl,floatfix,citeautoscript,nofootinbib]{revtex4-1}

\usepackage{amsbsy}
\usepackage{latexsym,epsfig,graphicx}
\usepackage{dcolumn}
\usepackage{graphicx}
\usepackage{subfigure}
\usepackage{comment}
\usepackage{color}
\usepackage[colorlinks,urlcolor=blue,citecolor=blue]{hyperref}

\begin{document}

\author{Junpeng Hou}
\author{Xi-Wang Luo}
\author{Kuei Sun}
\author{Chuanwei Zhang}
\thanks{Corresponding author. \\
Email: \href{mailto:chuanwei.zhang@utdallas.edu}{chuanwei.zhang@utdallas.edu}%
}
\affiliation{Department of Physics, The University of Texas at Dallas, Richardson, Texas
75080-3021, USA}
\title{Adiabatically tuning quantized supercurrents in an annular
Bose-Einstein condensate}

\begin{abstract}
The ability to generate and tune quantized persistent supercurrents is
crucial for building superconducting or atomtronic devices with novel
functionalities. In ultracold atoms, previous methods for generating
quantized supercurrents are generally based on dynamical processes to
prepare atoms in metastable excited states. Here we show that arbitrary
quantized circulation states can be adiabatically prepared and tuned as the
ground state of a ring-shaped Bose-Einstein condensate by utilizing
spin-orbital-angular-momentum (SOAM) coupling and an external
potential. There exists superfluid hysteresis for tuning supercurrents
between different quantization values with nonlinear atomic interactions,
which is explained by developing a nonlinear Landau-Zener theory. Our work
will provide a powerful platform for studying SOAM coupled ultracold atomic
gases and building novel atomtronic circuits.
\end{abstract}

\maketitle

\emph{Introduction}. Quantized supercurrents are one of the most remarkable
phenomena of superfluids and superconductors and have been widely studied in
solid-state superconductors~\cite{Tinkham2004} and ultracold atomic gases
\cite{Ryu2007,Corman2014}. Such persistent circulation currents are crucial
elements for building many important devices such as SQUID~\cite%
{Ryu2013,Sato2013,Clarke2004}, superfluid gyroscopes~\cite%
{Gustavson1997,Packard1992,Schwab1997} , atomic interferometers \cite%
{Kasevich1991,Mathew2015,Marti2015,Hoskinson2006,Simmonds2001} and
atomtronic circuits~\cite{Seaman2007,Pepino2009}. These important
applications naturally require the experimental ability of coherent
generation and manipulation of quantized supercurrents. In this context,
ultracold atomic gases possess intrinsic advantages for unprecedented
control of experimental parameters and the lack of disorder~\cite{Bloch2008}.

\begin{figure}[t]
\centering
\includegraphics[width=0.48\textwidth]{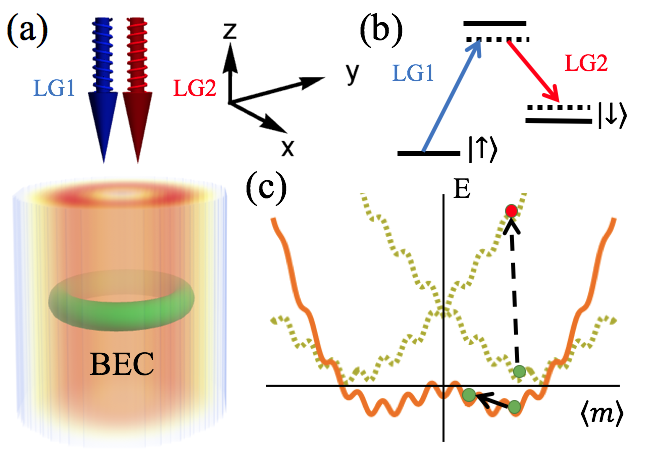}
\caption{(Color online) (a) Experimental scheme to generate a SOAM-coupled
ring BEC with co-propagating LG beams. (b) Raman transition induced by the
LG beams. (c) Schematic plot of the energy spectrum vs average quasi-OAM
number $m$ with and without Raman coupling (solid and dashed curves,
respectively). The energy barriers between integer $m$ are due to nonlinear
interactions. Green (red) circles denote ground (excited) states. Previous
experiments using a Raman short pulse dynamically excite the BEC to an upper
metastable state (dashed arrow) \protect\cite%
{Ryu2007,Moulder2012,Beattie2013}. While in our scheme the system remains on
the ground state during the adiabatic process (solid arrows).}
\label{fig1}
\end{figure}

Great progresses have been made recently for generating quantized
circulation supercurrents in a ring-shaped geometry and exploring these
properties and device applications~\cite%
{Ryu2007,Marzlin1997,Cooper2010,Moulder2012,Beattie2013,Ramanathan2011,Wright2013,Jendrzejewski2013,Eckel2014}%
. So far two experimental tools have been applied to prepare quantized
circulation currents for an annular Bose-Einstein condensate (BEC): \textit{i%
}) a short two-photon Raman pulse with orbital angular momentum (OAM)
transfer between two spin states \cite{Ryu2007,Moulder2012,Beattie2013}
using Laguerre-Gaussian (LG) laser beams~\cite{Allen1992}; \textit{ii})
periodic rotation of a local repulsive potential barrier along a ring \cite%
{Ramanathan2011,Wright2013,Jendrzejewski2013,Eckel2014}. These methods
involve dynamical process to transfer atoms to metastable high OAM states
[see Fig.~\ref{fig1} (c)], which could induce complicated excitations,
heating, and decays of the BEC. Therefore a nature question is whether
quantized supercurrents can be adiabatically prepared and manipulated as the
ground state of an annular BEC to circumvent these issues.

To prepare a finite OAM circulation state as the ground state, the OAM need
be coupled with other degrees of freedom such as spins, as we see from the
spin-orbit coupling \cite%
{Lin2011,Zhang2012b,Qu2013a,Olson2014,Hamner2014,Wang2012,Cheuk2012,Williams2013}%
. Recently, spin--OAM (SOAM) coupling has been proposed for ultracold atoms
using LG Raman lasers with finite OAM to couple two atomic hyperfine spin
states~\cite{Sun2015,DeMarco2015,Qu2015}. Since OAM is still a good quantum
number on a ring, an external nonuniform potential is needed to induce
coupling between different OAM states.

In this Letter, we show that the combination of these two ingredients, SOAM
coupling and nonuniform potential, allows adiabatic preparation and control
of quantized supercurrents with arbitrary OAM as the system ground state in
a ring-shaped BEC. Our main results are:

\textit{i}) A circulation state with arbitrary OAM can be generated on
demand with a high precision from a non-rotating state at a time scale $\sim
10\hbar /E_{R}$, where $E_{R}=\hbar ^{2}/2MR^{2}$ is the energy unit defined
by atomic mass $M$ and ring radius $R$. Large interaction strength or OAM
states shorten the time required for the adiabatic process. Such circulation
state carries both non-zero particle and spin supercurrents. Because the
adiabatic preparation process is free from complicated excitations and the
system is on ground state, the generated quantized supercurrents are stable
against heating and decaying of the BEC.

\textit{ii}) The tuning of supercurrents between different circulation
states strongly depends on the interatomic interaction and nonuniform
potential, which, in a proper parameter region, possesses superfluid
hysteresis.

\textit{iii}) We develop a nonlinear Landau-Zener theory to explain the
observed adiabatic preparation and superfluid hysteresis, which exhibits a
swallowtail energy structure due to nonlinear interactions.

%%%%%%%%%%%%%%%%%
%%End of Section - Introduction
%%%%%%%%%%%%%%%%%

\emph{Model and Hamiltonian}. We consider an atomic BEC with two internal
spin states $%
\begin{array}{cc}
{({\psi _{\uparrow }}} & {{\psi _{\downarrow }}{)^{T}}}%
\end{array}%
$, subject to a pair of co-propagating vertical LG beams with opposite OAM
number $\pm l$ [see Fig.~\ref{fig1}(a)] and $\Omega _{\pm }(r)=\Omega
_{0}\left( \frac{\sqrt{2}r}{w}\right) ^{|l|}\exp \left( -\frac{r^{2}}{w^{2}}%
\pm il\phi -ik_{L}z\right) $, where $(r,\phi ,z)$ are cylindrical
coordinates, and $\Omega _{0}$, $w$, and $k_{L}$ are the beams' amplitude,
waist, and wavevector in the $\hat{z}$ direction, respectively. The LG beams
induce a Raman transition [as in Fig.~\ref{fig1}(b)] that generates the SOAM
coupling \cite{Sun2015,DeMarco2015,Qu2015}. An additional far-detuned
vertical LG laser with the same beam waist and OAM provides the tube
potential around the maximum beam intensity $r=\sqrt{l/2}w$ and a horizontal
\textquotedblleft sheet\textquotedblright\ beam provides the confinement
along the $z$ direction \cite{Beattie2013}. Together the SOAM-coupled BEC is
confined on an annular geometry of a fixed radius $R$ and can be described
by an effective single-particle Hamiltonian~\cite{Sun2015} in units of $E_{R}
$ and $\hbar $ ($=1$), as
\begin{equation}
H_{0}^{\mathrm{ring}}=-\partial _{\phi }^{2}+\left( 2il\partial _{\phi
}+\delta /2\right) \sigma _{z}+\Omega \sigma _{x},
\end{equation}%
plus interatomic interactions. Here $\delta $ is the Zeeman detuning, $%
\Omega $ is the Raman coupling strength, and $\{\sigma \}$ are the Pauli
matrices.

The Hamiltonian has two energy bands. The eigenstates of the lower band are
plane waves in the $\hat{\phi}$ direction $\left\vert m\right\rangle ={(\cos
\theta _{m},-\sin \theta _{m}{)^{T}}{e^{im\phi }}}$, where $m$ is an integer
due to the periodic boundary condition. Note that $m$ represents a quasi-OAM
quantum, while the up and down components have real OAM $\propto (m\mp l)$~%
\cite{Sun2015}, respectively. In other words, the $\left\vert m\right\rangle
$ state physically carries quantized particle supercurrent $J_{c}(m)=m-l\cos
2\theta _{m}$ and spin supercurrent $J_{s}(m)=m\cos 2\theta _{m}-l$. The
ground-state supercurrent configurations change with $\Omega $. Given $%
\delta =0$ and $\Omega =0$, the lower band has two degenerate minima at $%
m=\pm l$ [as in Fig.~\ref{fig1}(c)]. As $\Omega $ increases, the double
minima shift toward each other and merge at $m=0$ when $\Omega =4l^{2}-1$,
above which the band has a single minimum $m=0$. In the double-minima
region, any pair of $\left\vert \pm m\right\rangle $ states exhibits a $Z_{2}
$ symmetry by having opposite spin polarization $\langle \sigma _{z}\rangle
=\cos 2\theta _{m}$, opposite $J_{c}$, and same $J_{s}$. Below we focus on
the plane-wave state on the $m\geq 0$ side.

\emph{Preparation of quantized supercurrent}. To create the ground state or
manipulate the supercurrent circulation in the ring system, one may properly
tune $\Omega $ to load the BEC at a desired quantum $m$, as it naturally
pursuits the energy minimum. However, there are two issues preventing this.
First, all $m$ states are stationary states and do not couple to each other,
so a high-$m$ state can have a very long lifetime. Second, a smooth
transition between adjacent $m$ states encounters an energy barrier induced
by interatomic interaction, which takes the form of
\begin{equation}
E_{g}=\frac{1}{2\pi }\int_{0}^{2\pi }\left( g_{\uparrow \uparrow }|\psi
_{\uparrow }|^{4}+g_{\downarrow \downarrow }|\psi _{\downarrow
}|^{4}+2g_{\uparrow \downarrow }|\psi _{\uparrow }|^{2}|\psi _{\downarrow
}|^{2}\right) d\phi .
\end{equation}%
In Fig.~\ref{fig1}(c), we schematically plot the system energy vs
expectation value of $m$ for a superposition of two adjacent $|m\rangle $
states. The ripples reflect the interaction effects: the local minima
correspond to pure $|m\rangle $ states, while the local maxima correspond to
the equal superposition of two adjacent $|m\rangle $ states, which possess
density modulations that cost interaction energy.

\begin{figure}[t]
\centering
\includegraphics[width=0.48\textwidth]{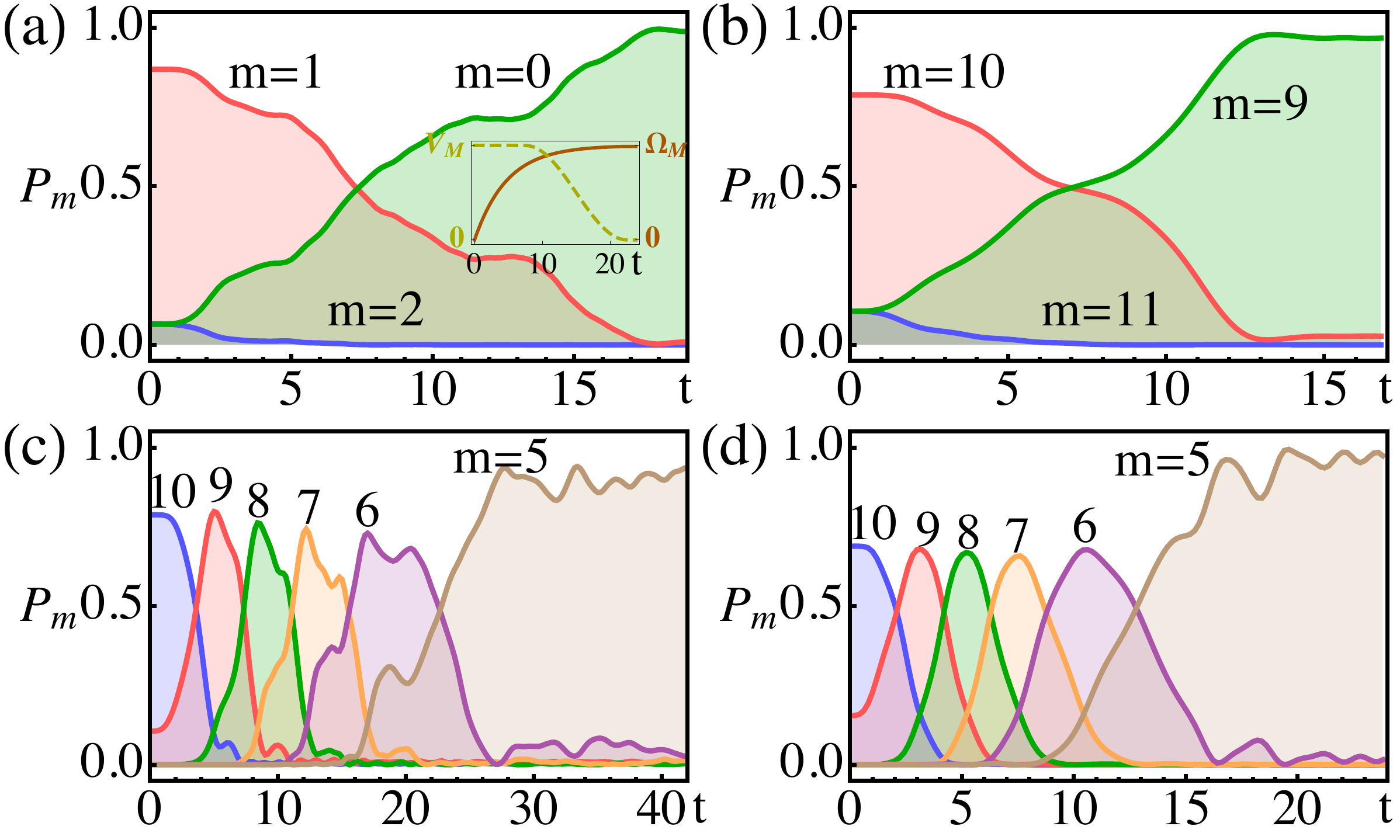}
\caption{(Color online) Time evolution of the BEC's population at $|m\rangle
$, or $P_{m}$, under the simultaneous increase in $\Omega $ and decrease in $%
V$ (solid and dashed curves in the inset, respectively). (a) Loading the BEC
to $|0\rangle $ from an initially prepared state with potential $V_{\mathrm{M%
}}=3$. The system parameters are $l=1$, $\Omega _{\mathrm{M}}=3$, $%
g_{\uparrow \uparrow }=15$, and $g_{\downarrow \downarrow }=g_{\uparrow
\downarrow }=0.9954g_{\uparrow \uparrow }$. (b) Loading to $|9\rangle $ with
higher LG beams $l=10$ and stronger $\Omega _{\mathrm{M}}=70$ in a similar
procedure. (c) Same initial state as (b), but the BEC transfers across
multiple states to $|5\rangle $ (with $\Omega _{\mathrm{M}}=180$). (d) Same
as (c) except stronger $g_{\uparrow \uparrow }=60$ and $V_{\mathrm{M}}=15$.}
\label{fig2}
\end{figure}

To assist the BEC to overcome the barrier and move toward a lower energy
state, it is essential to trigger sufficient coupling between adjacent $%
|m\rangle $ states. We propose the use of a linear external potential $V_{%
\mathrm{ext}}=V(t)x/R=V(t)\cos \phi $, which couples two states as $\langle
m|\cos \phi |m\pm 1\rangle =\frac{1}{2}\langle m|e^{i\phi }+e^{-i\phi }|m\pm
1\rangle \neq 0$. Such potential has already been experimentally applied to
annular BECs without SOAM coupling~\cite{Kumar2016}. In addition, one can
use the linear slope around $x=0$ for a Gaussian laser $V\left( x\right)
=V_{0}\exp \left( -(x-\xi )^{2}/2\xi ^{2}\right) \approx V_{0}\exp \left(
-1/2\right) \left( 1+x/\xi \right) $ with $\xi \gg R$. We propose a process
of tuning both $\Omega $ and $V$ simultaneously, as illustrated in the inset
of Fig.~\ref{fig2}(a), and simulate the system evolution with the
time-dependent Gross-Pitaevskii equation (GPE). Initially, the BEC is
prepared in the presence of external potential $V(0)=V_{\mathrm{M}}$. The
Raman coupling $\Omega $ is then slowly ramped on to shift the band minimum,
while $V$ is gradually turned off to suppress the coupling. Finally, at $%
\Omega (t)=\Omega _{\mathrm{M}}$ and $V(t)=0$, the BEC stays in the new
ground state, the targeted single-$m$ state, which decouples from the others
and carries the aforementioned particle and spin supercurrents.

Our simulation of four different cases is presented in Fig.~\ref{fig2}. We
plot the BEC's population $P_{m}$ at several associated $m$ states as a
function of time. In panel (a) for $l=1$, the BEC starts from a $|1\rangle $
dominant state and ends at an almost single-$m$ state, $|0\rangle $,
demonstrating an effective transfer between adjacent states. Although the
initial state also couples to $|2\rangle $, the BEC clearly pursuits the
lower energy state $|0\rangle $. In general, higher energy states are hardly
involved during the transition. The same procedure can be applied to a large
$l=10$, as shown in Fig.~\ref{fig2}(b), where only $\Omega _{\mathrm{M}}$ is
changed for the desired band minimum. The system actually undergoes a
smoother transition with a shorter transition time for a large $l$. Using
the same strategy, one may possibly reach any $m$ state ($0\leq m\leq l$) by
repeating the $\Omega $ and $V$ cycle to lower the $m$ number one by one.
However, such a transfer, e.g.~from $|10\rangle $ to $|5\rangle $, can be
achieved with a single process and hence in a shorter time [see Fig.~\ref%
{fig2}(c)]. The BEC passes through multiple $m$ states with each
intermediate $m$ state dominating in a narrow time window, and stops at the
final $m$ state. In Fig.~\ref{fig2}(d), we quadruple the interatomic
interaction and increase $V_{\mathrm{M}}$ accordingly to overcome the
interaction energy barrier. The transition goes more smoothly and the
transition time is significantly shorter.

%%%%%%%%%%%%%%%%%%%%%%%%%%%
%%End of Section - Prepare arbitrary quantized state
%%%%%%%%%%%%%%%%%%%%%%%%%%%

\begin{figure}[t]
\centering
\includegraphics[width=0.48\textwidth]{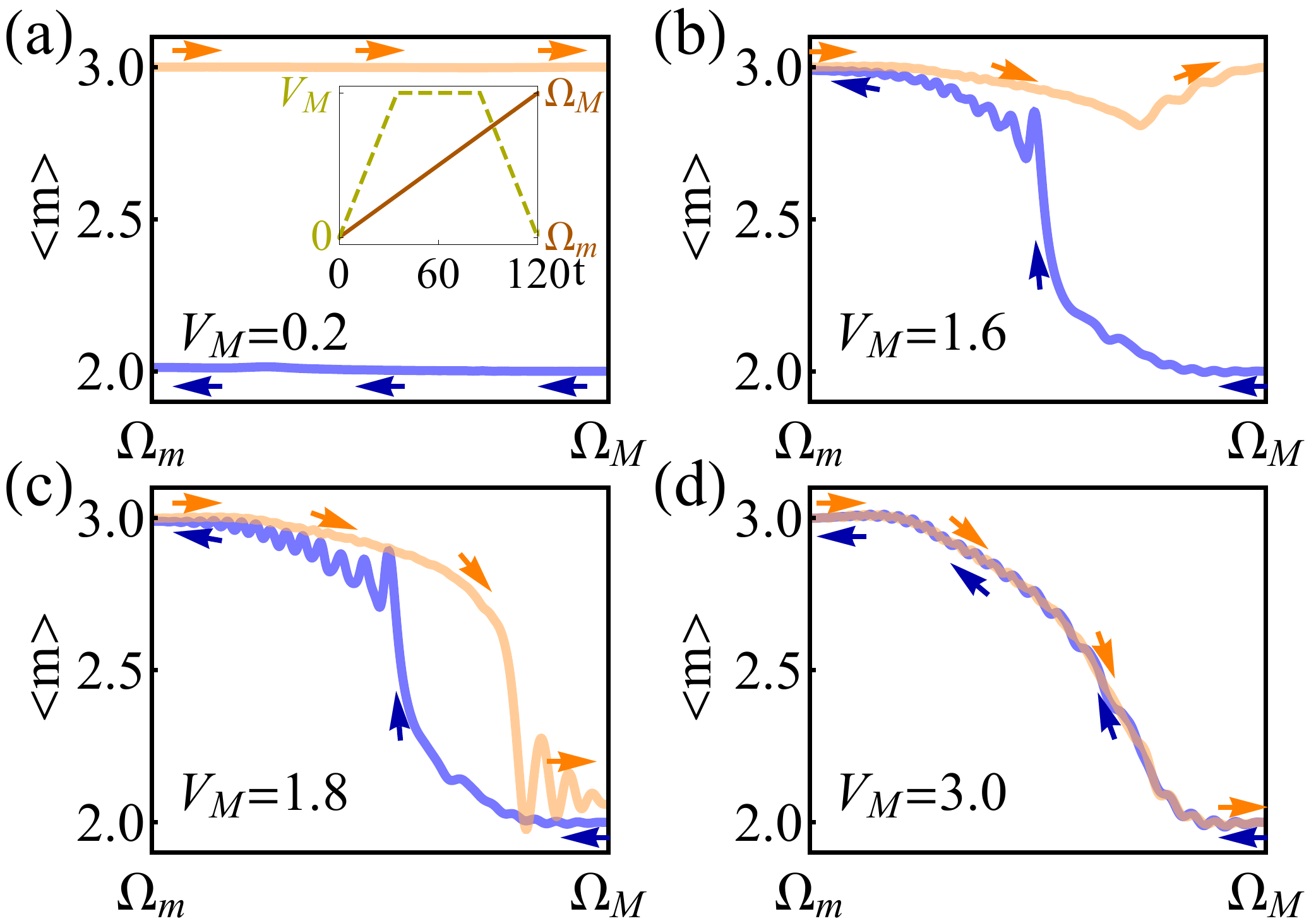}
\caption{(Color online) Inner panel shows time procedure of $\Omega (t)$
(solid curve, axis on the right) and $V(t)$ (dashed curve, axis on the left)
for studying the superfluid hysteresis. The Raman coupling $\Omega _{\mathrm{%
m}}$ ($\Omega _{\mathrm{M}}$) corresponds to the ground state being $%
|3\rangle $ ($|2\rangle $). (a)--(d) Time evolution of the expectation value
$\langle m\rangle $ shows no transition at $V_{\mathrm{M}}=0.2$, single-side
transition at $V_{\mathrm{M}}=1.6$, a hysteretic loop at $V_{\mathrm{M}}=1.8$%
, and a loop without hysteresis at $V_{\mathrm{M}}=3$, respectively. The
forward evolution (orange curve) starts from $|3\rangle $, and the reversal
(blue curve) starts from $|2\rangle $, with both directions indicated by the
arrows.}
\label{fig3}
\end{figure}

\emph{Superfluid hysteresis}. Hysteresis is a hallmark phenomena of
quantized supercurrents \cite{Pepino2002,Watanabe2011,Morsch2006}, and
recently a hysteresis loop for supercurrent jumps in an annular BEC has been
observed in experiments, where a periodically rotating local laser barrier
is used to induce the transition~\cite{Eckel2014}. In our system, the
supercurrent change is driven by the SOAM coupling to change the ground
state, which, unlike the local laser barrier, acts on the whole system
uniformly. Below we will show the existence of superfluid hysteresis using
numerical GPE simulation (Fig.~\ref{fig3}), followed by developing a
nonlinear Landau-Zener theory~\cite{Vitanov1999,Liu2003,Wu2000} to explain
the physics, where a swallowtail band structure is ascribed to the origin of
the observed hysteretic phenomenon.

The time procedure of $\Omega (t)$ and $V(t)$ in our simulation is
illustrated in Fig.~\ref{fig3}(a) inset. In the forward evolution, $\Omega
(t)$ linearly increases from $\Omega _{\mathrm{m}}$ to $\Omega _{\mathrm{M}}$
with the initial ground state prepared at $\Omega _{\mathrm{m}}$, while $V(t)
$ is turned on up to the maximum $V_{\mathrm{M}}$ and then symmetrically
turned off. In the backward evolution, $\Omega (t)$ and $V(t)$ take the
time-reversal path, with the initial ground state prepared at $\Omega _{%
\mathrm{M}}$. Since $V(t)$ is the key ingredient that induces the
transition, we expect that it plays a crucial role on the hysteresis
phenomenon. In Fig.~\ref{fig3}(a)--(d), we plot the expectation value $%
\langle m\rangle $ for transitions between $|2\rangle $ and $|3\rangle $ at
various $V_{\mathrm{M}}$. At weak $V_{\mathrm{M}}=0.2$, there is no
transition [Fig.~\ref{fig3}(a)]. At $V_{\mathrm{M}}=1.6$ [Fig.~\ref{fig3}%
(b)], the system cannot jump from $|3\rangle $ to $|2\rangle $ before the
coupling is turned off, leaving the path unclosed. With the increase of the
potential to $V_{\mathrm{M}}=1.8$ [Fig.~\ref{fig3}(c)], a two-way transition
happens with a clear hysteresis loop. At a strong $V_{\mathrm{M}}=3$ [Fig.~%
\ref{fig3}(d)], the coupling completely overwhelms the barrier, and the loop
becomes trivial, as the system follows the same path forward and backward.

\begin{figure}[t]
\centering
\includegraphics[width=0.48\textwidth]{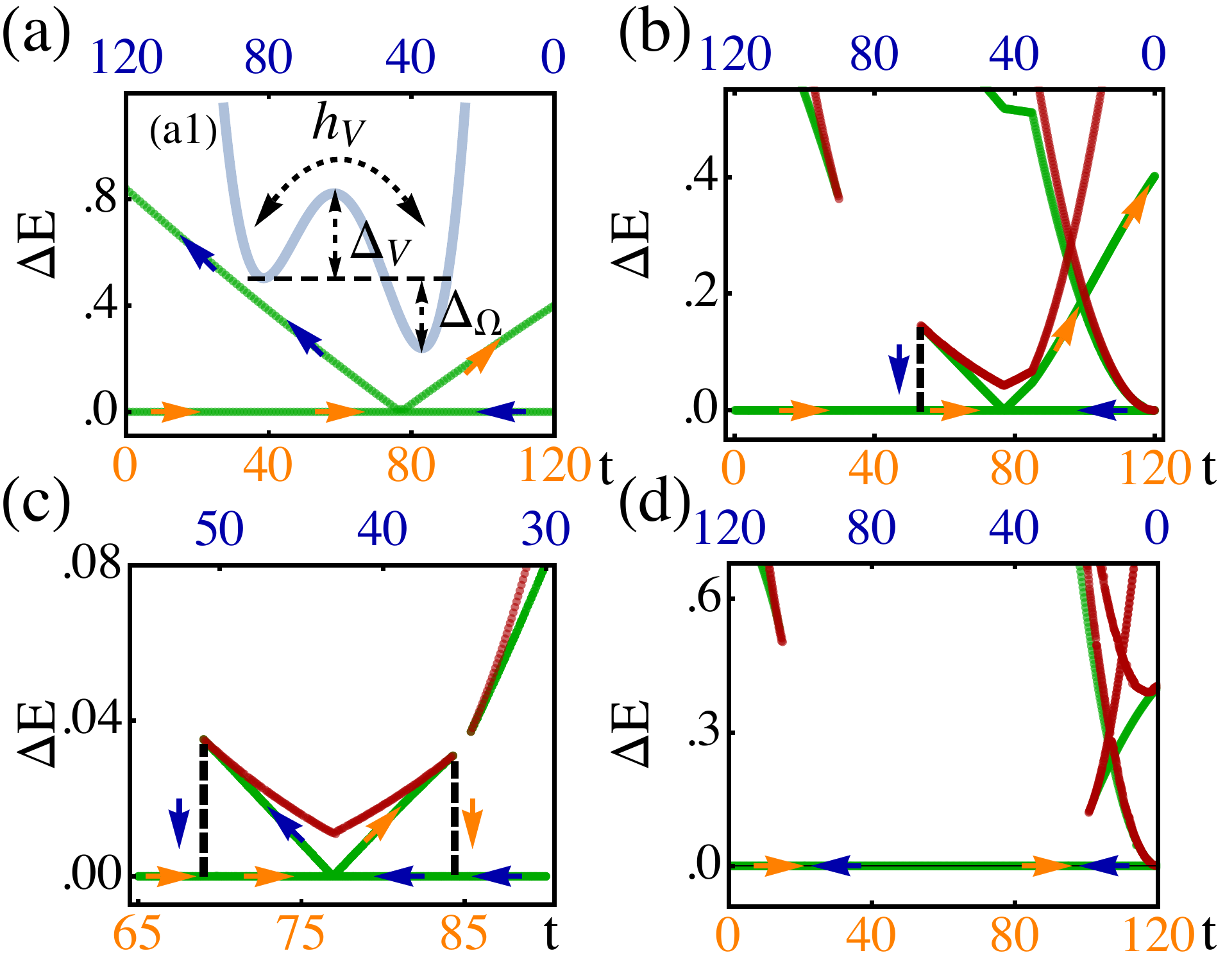}
\caption{(color online) (a1) Schematic plot of the double-well structure of
the system in the phase space, characterized by the energy difference
between the two minima $\Delta _{\Omega }$, barrier height $\Delta _{V}$,
and hopping strength $h_{V}$. (a)--(d) Energy band structure during the
evolution in Fig.~\protect\ref{fig3}(a)--(d), respectively. The green and
red curves represent the band minima and maxima (or saddle points),
respectively. The arrows indicate the evolution direction, same as in Fig.~%
\protect\ref{fig3}, and the dashed lines indicate the sudden jump to ground
states. Panel (c) is rescaled to emphasize the swallow-tail structure, which
explains how the transitions happen differently between the forward and
backward evolutions, resulting in the superfluid hysteresis.}
\label{fig4}
\end{figure}

The above observed phenomena can be intuitively illustrated from a double
well energy structure [Fig.~\ref{fig4}(a1)] in proper phase space of
parameters of the system. Initially, BEC always stays in one of the minimum
(ground state). During the evolution, the energy difference $\Delta _{\Omega
}$ between the two minima changes with $\Omega (t)$, while the barrier
height $\Delta _{V}$ is determined by the interaction and $V(t)$. If $V_{%
\mathrm{M}}$ is sufficiently weak, the two minima are well separated by the
barrier, and the hopping $h_{V}$ is not strong enough to overcome it. As a
result, the BEC is trapped in one minimum without any transition. As $V_{%
\mathrm{M}}$ increases, the minimum on which the BEC originally stays
disappears during the process, and the BEC jumps to the other minimum. Such
evolution is not symmetric in the time-reversal path, leading to the
hysteretic behavior [Fig.~\ref{fig3}(c)]. If $V_{\mathrm{M}}$ is strong, the
initial minimum continuously evolves to the final minimum, and vice versa,
resulting in the trivial loop [Fig.~\ref{fig3}(d)].

This intuitive picture is confirmed by developing a nonlinear Landau-Zener
theory for our system. As shown in Fig. \ref{fig2}, the transition process
between $|m\rangle $ and $|m-1\rangle $ barely involves other higher-energy
states, which allows us to build an effective Hamiltonian in a truncated
space spanned by two relevant quasi-OAM and two spin states. The system's
wavefunction is determined by four complex amplitudes as $\left(
\begin{array}{c}
a_{1} \\
a_{2}%
\end{array}%
\right) {{e^{i(m-1)\phi }+}}\left(
\begin{array}{c}
a_{3} \\
a_{4}%
\end{array}%
\right) {{e^{im\phi }}}$, subject to a normalization condition $%
\sum_{j=1}^{4}p_{j}=1$ with $p_{j}=|a_{j}|^{2}$.

Following the general formalism for nonlinear Landau-Zener tunneling~\cite%
{Liu2003,Wu2000}, we construct a semi-classical Hamiltonian with canonical
coordinates ($q_{j}=\arg a_{j}$, $p_{j}$)
\begin{eqnarray}
&&\mathcal{H}=(l-m+1)^{2}p_{1}+(l-m)^{2}p_{3}+(l+m-1)^{2}p_{2} +  \nonumber
\label{MiniH} \\
&&(l+m)^{2}p_{4}+2V[\sqrt{p_{1}p_{3}}\cos (q_{1}-q_{3})+\sqrt{p_{2}p_{4}}%
\cos q_{2}] +  \nonumber \\
&&2\Omega \lbrack \sqrt{p_{1}p_{2}}\cos (q_{1}-q_{2})+\sqrt{p_{3}p_{4}}\cos
q_{3}] +  \nonumber \\
&&(g/\pi )[p_{1}p_{3}+p_{2}p_{4}+\sqrt{p_{1}p_{2}p_{3}p_{4}}\cos
(q_{1}-q_{2}-q_{3})],
\end{eqnarray}%
with a global phase chosen for $q_{4}=0$ and $g=g_{\uparrow \uparrow
}=g_{\downarrow \downarrow }=g_{\downarrow \uparrow }$.

For each $\Omega \left( t\right) $ and $V\left( t\right) $ during the
transition, we can find a global energy minimum (ground-state energy) in the
phase space formed by six parameters $\left( p_{i},q_{i}\right) $ ($i=1,2,3$%
). There also exist other interaction-induced local energy minima, as shown
in the energy structure in Fig.~\ref{fig4}(a)--(d), with parameters
corresponding to Fig.~\ref{fig3}(a)--(d), respectively. In particular, a
double-well structure with the emergence of two energy minima similar as
Fig.~\ref{fig4}(a1) is observed in the phase space (see Supplementary
Materials), which confirms the above intuitive explanation of the hysteretic
behavior. Because of multiple local energy minima for the same parameter,
the energy band structure in Fig.~\ref{fig4} shows swallowtail shape in
proper parameter region. In Fig.~\ref{fig4}(a), two bands simply cross and
both minima survive during the procedure. As a result, the condensate
remains on its initial minimum, which ends as a metastable state, without
undergoing a transition [Fig.~\ref{fig3}(a)].
With larger potential [Fig.~\ref{fig4}(b)], only the left-hand-side band
minimum breaks. Therefore, when starting from $|2\rangle $, the minimum
disappears after the cross, forcing BEC to drop to the ground state. This
result agrees with the GPE simulation in Fig.~\ref{fig3}(b). With a deeper
trap, the breakdown of band minimum is symmetric [Fig.~\ref{fig4}(c)],
therefore abrupt changes in $\langle m\rangle $ happen in both directions,
yielding the hysteresis loop observed in Fig.~\ref{fig3}(c). After $V_{%
\mathrm{M}}$ exceeds some critical value, the barrier vanishes and the BEC
always stays on the ground state [Fig.~\ref{fig4}(d)]. Therefore the system
follows exactly the same path during the evolution as observed in Fig.~\ref%
{fig3}(d). Our quantized supercurrent preparation should use this parameter
region.
%%%%%%%%%%%%%%%%%%%%%%%%%%
%%End of Section - Hysteresis and band structure
%%%%%%%%%%%%%%%%%%%%%%%%%%

\emph{Experimental consideration}. In experiments, we can consider a ${}^{87}
$Rb gas~\cite{Ramanathan2011,Wright2013} trapped in a ring of radius $%
R=8~\mu $m, and use LG beams $l=10$~\cite{Moulder2012}. The energy unit $%
E_{R}=2\pi \hbar \times 0.924$ Hz. For the transition procedure ($|10\rangle
$ to $|5\rangle $) shown in Fig.~\ref{fig2} (d), $\Omega _{M}=2\pi \hbar
\times 166$ Hz, $V_{M}=2\pi \hbar \times 14$ Hz and the overall time length $%
T=3.4$ s. For a small Raman coupling strength $\Omega _{M}$ around $1$ kHz,
the heating effect can be neglected within a timescale of few seconds~\cite%
{Wei2013}. Considering the same configuration but with ${}^{23}$Na~\cite%
{Stenger1998}, the parameters become $E_{R}=2\pi \hbar \times 3.434$ Hz, $%
\Omega _{M}=2\pi \hbar \times 618$ Hz, $V_{M}=2\pi \hbar \times 52$ Hz, $%
T=0.9$ s.

%%%%%%%%%%%%%%%%%%%%%%
%%End of Section - Experimental aspects
%%%%%%%%%%%%%%%%%%%%%%

\emph{Conclusion}. We have proposed a method to adiabatically prepare and
tune arbitrary quantized circulation states as the ground states of an
annular BEC, which carry both quantized atom and spin supercurrents. The
whole procedure can be achieved at a time scale $\sim 10\hbar /E_{R}$ with a
satisfactory fidelity. Our system provides a powerful platform for studying
phenomena and applications of quantized supercurrents, exploring physics of
SOAM coupled atomic gases, and building atomtronic devices with novel
functionalities.

%%%%%%%%%%%%%%%%
%%End of Section - Conclusion
%%%%%%%%%%%%%%%%

\textbf{Acknowledgements}: This work is supported by AFOSR
(FA9550-16-1-0387), NSF (PHY-1505496), and ARO (W911NF-17-1-0128).

%%%%%%%%%%%%%%%%%
%%End of Section - Acknowledge
%%%%%%%%%%%%%%%%%

\newpage \clearpage

\begin{appendix}
\section{Supplementary Materials}
\subsection{Time scale of the adiabatic process}
\begin{figure}[t]
    \centering
    \includegraphics[width=0.5\textwidth]{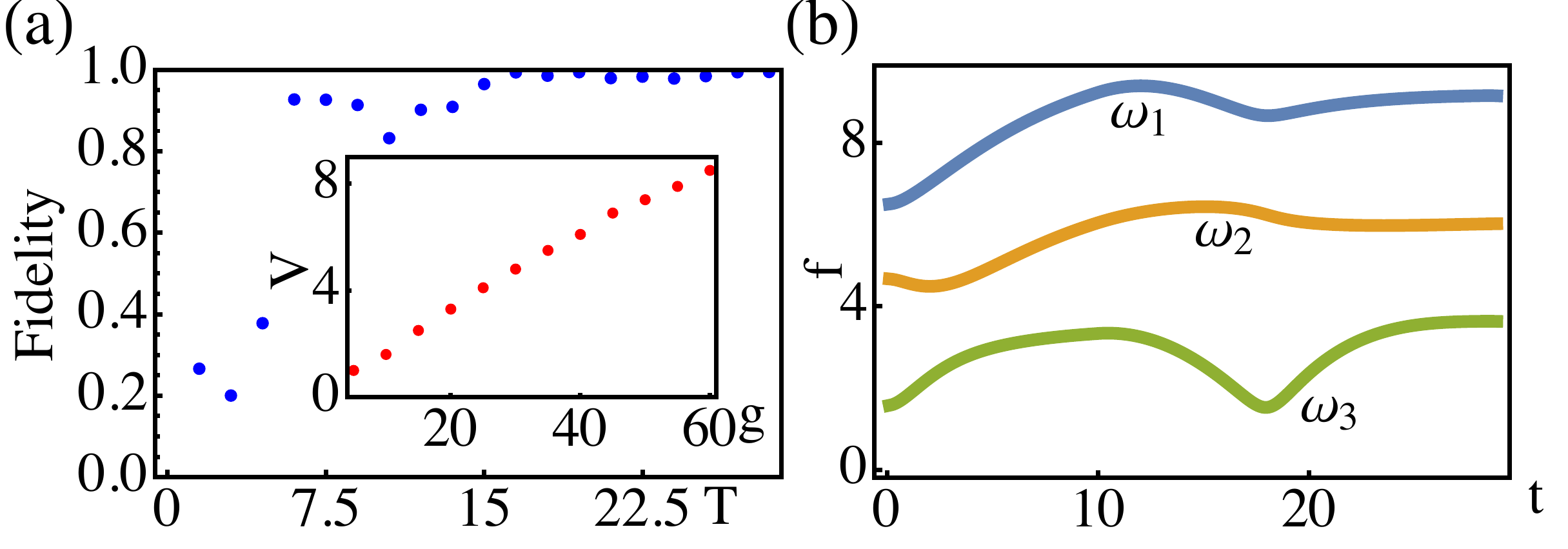}
    \caption{(color online) (a) Final state fidelity versus the overall time length.
    The inner panel gives
    the minimum $V_{\rm{M}}$ required to induce the smooth transition for different interaction strength.
    (b) Fundamental frequencies of the periodic orbits at the energy minimum.
    For both plots, unmentioned parameters are the same as the transition process in Fig.~\ref{fig2} (a).}
    \label{figs1}
\end{figure}

For the adiabatic process
given in Fig.~\ref{fig2} (a),
$V_{\rm{M}}$ should be strong enough to overcome the energy barrier
between different $|m \rangle$ states induced by the interaction in order to obtain the desired final state.
Therefore a larger $V_{\rm{M}}$ is required for stronger interaction with a rough linear relation
as shown in the inset in Fig.~\ref{figs1} (a).
The red dots show the minimum values of $V_{\rm{M}}$ to achieve a smooth transition between different $|m \rangle$ states,
where the energy minima for
adjacent $|m \rangle$ states are smoothly
connected by the adiabatic path.

The required time length of the adiabatic process
is closely related to the choice of the
adiabatic path in the parameter space, which is determined by
$V_{\rm{M}}$ and $\Omega_{\rm{M}}$.
The value of $\Omega_{\rm{M}}$ can be simply
determined by examining the single-particle band structure (or phase diagram).
As discussed above, for a given interaction strength,
a perfect transition can be realized by slowly tuning the parameters along the path with a large $V_{\rm{M}}$.
If the overall time length $T$ is too small, excitations may be created during the process,
and consequently, imperfections are introduced in the transition.
Fig.~\ref{figs1} (a) illustrates the relations between fidelity and overall time length ($T$) of the process
in Fig.~\ref{fig2} (a).
The fidelity is defined as the modulus square of the overlap between the final state
and the target state.
As $T$ decreases, excitations and their interference
with the ground state lead to the fidelity oscillation near $T=7.5$, and
the adiabatic approach breaks down below the time length $T=7$, where a sudden drop of the fidelity occurs.

For nonlinear quantum problems, the adiabaticity is related to
not only the energy-level spacings, but also the fundamental frequencies of
periodic orbits around the energy minima.
These frequencies can be calculated by linearizing the
semi-classical Hamiltonian Eq.~\ref{MiniH} around the energy minima.
One should expect to have three pairs of frequencies ($\omega_1,\omega_2,\omega_3$),
since the degrees of freedom $6$ (i.e. the dimension of the
phase space is 6).
Fig.~\ref{figs1} (b) shows how these fundamental frequencies
vary with time during the adiabatic process shown in Fig.~\ref{fig2} (a).
The minimum frequency is about 1,
indicating that the overall time length should be much longer than
1 to avoid excitations of such periodic orbits.
For the processes shown in
Fig.~\ref{fig2} (a) and Fig.~\ref{fig3} (d),
the energy minimum is far away from any energy
maxima in the phase space, therefore the time scales of the adiabatic processes are determined by
the fundamental frequencies, as confirmed by the results shown in
Fig.~\ref{figs1} (a).
For the hysteresis process, the minimum
merges with a maximum and both of them disappear after the merging, therefore the periodic orbits
may not even exist and these fundamental frequencies lose their physical
meanings.
In fact, the adiabaticity breaks down in the hysteresis process.
Nevertheless, our simulation suggests that a longer time is always preferred
to get a smoother hysteresis loop.

\begin{figure}[t]
    \centering
    \includegraphics[width=0.5\textwidth]{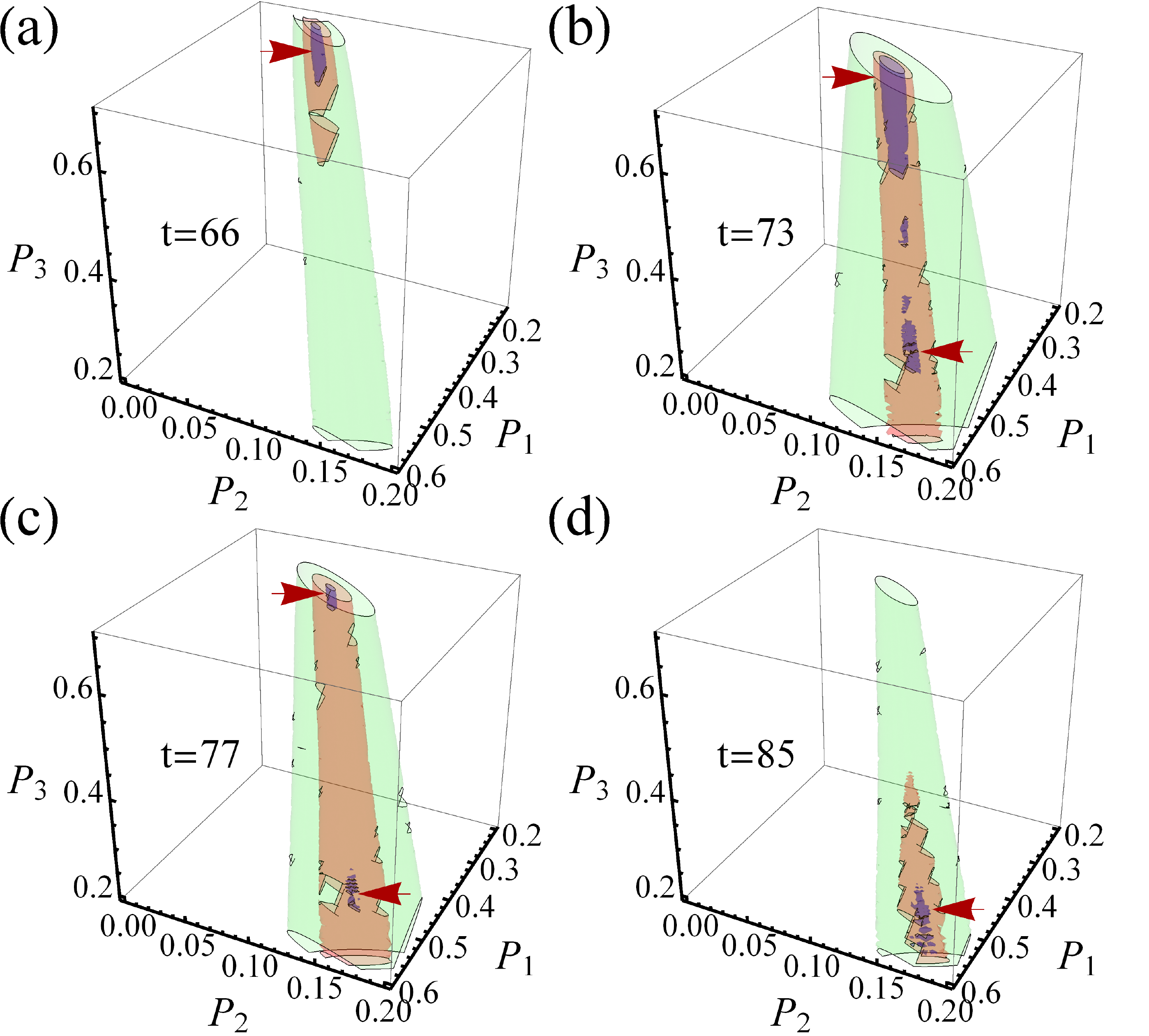}
    \caption{(color online) Phase space portraits in the subspace spanned by $(p_1,p_2,p_3)$ with
    $q_1=2\pi$, $q_2=\pi$ and $q_3=\pi$.
    Each panel contains three equal-energy surfaces that increase from inner to outside, they are given by $(-10.41,-10.38,-10.3)$ in (a), $(-10.83,-10.8,-10.5)$ in (b), $(-11.1,-11.05,-10.9)$ in (c), and  $(-11.64,-11.62,-11.55)$ in (d). The red arrows denote the approximate positions of the minima.}
    \label{figs2}
\end{figure}

\subsection{Energy minima in the phase space}
The evolution of energy minima in the phase space
are crucial for understanding the
adiabatic process. As an example, we consider the hysteresis process discussed in Fig.~\ref{fig3} (c).
To visualize the energy minima in the
phase space with dimension as high as $6$,
we first notice that the minima always satisfy
$q_1=2\pi$, $q_2=\pi$ and $q_3=\pi$.
This gives
us the possibility to project the whole phase space to a 3D subspace spanned by $(p_1,p_2,p_3)$.
At $t=66$, there is only one minimum that gives
the ground state, as shown in Fig.~\ref{figs2} (a).
As we change $\Omega$ and $V$, a second local minimum appears [see Fig.~\ref{figs2} (b) at $t=73$]. This process is accompanied by the emergence of a
local maximum that forms the barrier between the two minima.
Though the second minimum may have a lower energy, the BEC will stay at the first minimum
[see Fig.~\ref{figs2} (c) at $t=77$] until it merges with
the maximum and disappears, after which the
BEC flows towards the second minimum [see Fig.~\ref{figs2} (d) at $t=85$].
Those phase-space portraits shown in Fig.~\ref{figs2} provide us
a clear picture of the physics behind the swallowtail band structure and the corresponding hysteresis. They confirm our intuitive explanation of the hysteresis phenomena illustrated in Fig.~\ref{fig4}(a).
%%%%%%%%%%%%%%%%%%%%%%%%%
%%End of Appendix. Section - General Transition
%%%%%%%%%%%%%%%%%%%%%%%%%

\end{appendix}

\end{document}